# Simple forms of SLOCC equivalent four-qubit $\chi$ state

Xin-Wei Zha, Jian-Xia Qi and Yun-Guang Zhang

School of Science, Xi'an University of Posts and Telecommunications, Xi'an, 710121,P R China。

**Abstract**   Some simple structure of maximally four-qubit state is presented. Using stochastic local operations and classical communication (SLOCC) invariants, it is can be shown those simple structure of maximally four-qubit state is SLOCC equivalent to four-qubit $\chi$ state.

## 1 Introduction

Quantum entanglement is a valuable resource for the implementation of quantum computation and quantum communication protocols, like quantum teleportation [1], remote state preparation[2] and so on. Due to its great relevance, from both the theoretical and practical points of view, it is imperative to explore and characterize all aspects of the quantum entanglement of multipartite quantum systems. A common feature of entanglement for multipartite quantum systems is that the non-local properties do not change under local transformations, i.e., the unitary operations act independently on each of the subsystems. There are a series of works concentrated on revealing the relationship between SLOCC invariant and entanglement.[3–12].The SLOCC equivalence was introduced in [5] and enjoys a clear physical motivation。 two multipartite pure states $|\psi\rangle$ and $|\varphi\rangle$ are equivalent under SLOCC if one can be obtained from the other with non-null probability using only local operations upon each subsystem and classical communication among the different parties.

On the other hand，Yeo and Chua [13] presented maximally four-qubit state。As one of important entangled states, $\chi$-type entangled state has many interesting properties and was used to implement quantum key distribution, quantum secret sharing, quantum secure direct communication, and so on [13–14].

In this paper, we introduce some simple structure of maximally four-qubit state，which is SLOCC equivalent to maximally four-qubit entangled $\chi$ states。

## 2. Simple structure of maximally four-qubit state

The states of a four-qubit system can be generally expressed as

$$|\psi\rangle_{ABCD} = a_0|0000\rangle + a_1|0001\rangle + a_2|0010\rangle + a_3|0011\rangle$$
$$+ a_4|0100\rangle + a_5|0101\rangle + a_6|0110\rangle + a_7|0111\rangle$$
$$+ a_8|1000\rangle + a_9|1001\rangle + a_{10}|1010\rangle + a_{11}|1011\rangle$$
$$+ a_{12}|1100\rangle + a_{13}|1101\rangle + a_{14}|1110\rangle + a_{15}|1111\rangle \quad (1)$$

Two states $|\psi\rangle$ and $|\varphi\rangle$ are equivalent under SLOCC if and only if there exist invertible local operators $M_A, M_B, M_C \cdots$, and such that

$$|\psi\rangle_{ABC\cdots} = M_A \otimes M_B \otimes \cdots \otimes |\varphi\rangle_{ABC\cdots} \quad (2)$$

where the local operators $M_A, M_B, M_C \cdots$, and can be expressed as $2 \times 2$ invertible matrices。

It is know there are four independent of invariants, $(H, L, M, D_{xt})$ for a 4-qubit system[9]. H, a degree-2 invariant can be directly written as

$$H = 2(a_0 a_{15} - a_1 a_{14} - a_2 a_{13} + a_3 a_{12} - a_4 a_{11} + a_5 a_{10} + a_6 a_9 - a_7 a_8) \quad (3)$$

The two degree-4 invariants L,M, are given by the determinants of three matrices:

$$L = \begin{vmatrix} a_0 & a_4 & a_8 & a_{12} \\ a_1 & a_5 & a_9 & a_{13} \\ a_2 & a_6 & a_{10} & a_{14} \\ a_3 & a_7 & a_{11} & a_{15} \end{vmatrix} \quad (4)$$

$$M = \begin{vmatrix} a_0 & a_8 & a_2 & a_{10} \\ a_1 & a_9 & a_3 & a_{11} \\ a_4 & a_{12} & a_6 & a_{14} \\ a_5 & a_{13} & a_7 & a_{15} \end{vmatrix} \quad (5)$$

the $D_{xt}$ is a degree-6 invariants[11], and can be expressed as the determinants of three $3 \times 3$ matrices:

$$D_{xt} = \begin{vmatrix} a_0a_6-a_2a_4, & a_0a_7+a_1a_6-a_2a_5-a_3a_4, & a_1a_7-a_3a_5, \\ a_0a_{14}+a_6a_8 & a_0a_{15}+a_6a_9+a_1a_{14}+a_7a_8 & a_1a_{15}+a_7a_9 \\ -a_2a_{12}-a_4a_{10}, & -a_2a_{13}-a_3a_{12}-a_4a_{11}-a_5a_{10}, & -a_3a_{13}-a_5a_{11}, \\ a_8a_{14}-a_{10}a_{12}, & a_8a_{15}+a_9a_{14}-a_{10}a_{13}-a_{11}a_{12}, & a_9a_{15}-a_{11}a_{13}, \end{vmatrix} \quad (6)$$

In ref[15], we have introduced 16 different SLOCC entanglement classes via the the independent SLOCC invariant for four qubits. And that four-qubit entangled $\chi$ state [12]:

$$|\chi\rangle_{1234} = \frac{1}{2\sqrt{2}}(|0000\rangle - |0011\rangle - |0101\rangle + |0110\rangle + |1001\rangle + |1010\rangle + |1100\rangle + |1111\rangle)$$

It is easy to show that $H=0$, $L=-\frac{1}{16}$, $M=\frac{1}{16}$, $D_{xt}=0$.

Now let us analysis state

$$\left|\varphi_M^1\right\rangle_{1234} = \frac{1}{2}(|0000\rangle + |0111\rangle - |1001\rangle + |1110\rangle)_{1234} \quad (7)$$

i.e., $a_0 = a_7 = a_{14} = \frac{1}{2}$, $a_9 = -\frac{1}{2}$

It is easy to show that $H=0$, $L=-\frac{1}{16}$, $M=\frac{1}{16}$, $D_{xt}=0$.

That is to say, the state $\left|\varphi_M^1\right\rangle_{1234}$ is SLOCC equivalent to the state $|\chi\rangle_{1234}$

### 3. Discussions and conclusions

Also, we can find that the state

$$\left|\varphi_M^2\right\rangle_{1234} = \frac{1}{2}(|0000\rangle - |0111\rangle + |1001\rangle + |1110\rangle)_{1234}$$

is also SLOCC equivalent to the state $|\chi\rangle_{1234}$。

Furthermore, we find

$$\left|\varphi_M^1\right\rangle_{1234} = H_1 \otimes H_2 \otimes H_3 |\chi\rangle_{1234}$$

$$\left|\varphi_M^2\right\rangle_{1234} = \hat{\sigma}_{1x} \otimes H_2 \otimes H_3 \otimes H_4 \left|\chi\right\rangle_{1234}$$

Where $\hat{\sigma}_{1x}$ is Pauli operator and $H_i$ is the Hadamard transformation。

Therefore，$\left|\varphi_M^1\right\rangle_{1234}$ and $\left|\varphi_M^2\right\rangle_{1234}$ are also maximally entangled four-qubit states. In summary, we have introduced new forms SLOCC equivalent to the $\chi$ state 。we have shown that there exists a relation between the simple forms state and the $\chi$ state。It is will know that $\chi$-type entangled state has many interesting properties and was used to implement many quantum information work。Therefore，we believe ours simple forms state will be useful in the future。

Acknowledgements

This work is supported by the Shaanxi Natural Science Foundation under Contract (No. 2013JM1009).